\documentclass[11pt]{article}

\usepackage[a4paper,margin=1in]{geometry}
\usepackage{amsmath,amssymb,amsthm}
\usepackage{enumitem}
\usepackage{booktabs}
\usepackage{xcolor}
\usepackage{url}
\usepackage{hyperref}
\definecolor{linkblue}{RGB}{30,80,180}
\definecolor{linkgreen}{RGB}{30,120,60}
\definecolor{linkred}{HTML}{AE0000}

\hypersetup{
  colorlinks=true,
  linkcolor=linkgreen,
  citecolor=linkgreen,
  urlcolor=linkblue,
  filecolor=linkred
}

\title{\textbf{The Ghost in the Datacenter:}\\[0.3em]
{\large Link Flapping, Topology Knowledge Failures,\\
and the FITO Category Mistake}}

\author{
Paul Borrill \\
D{\AE}D{\AE}LUS \\
\texttt{paul@daedaelus.com} \\
ORCID: 0000-0002-7493-5189
}

\date{March 2026 \quad$\vert$\quad Revised September 2026 (v2.4)}

\newtheorem{definition}{Definition}

\newtheorem{proposition}{Proposition}
\newtheorem{corollary}{Corollary}

\begin{document}

\maketitle

\begin{center}
\fbox{\parbox{0.92\textwidth}{\small
\textbf{Changes in v2 (September 2026).}  This revision follows a
self-audit against the OAE specification and two adversarial reviews.
\textbf{It weakens claims and corrects errors; it adds no new results.}

\emph{Corrections.}  A TPUv4 availability figure was misconverted
(99.98\% is ${\approx}105$~minutes per pod-year, not 18) \emph{and}
wrongly equated with ghost duration; both are fixed.
Proposition~\ref{prop:tar} previously asserted an upper bound on the
ghost window of ``timeout plus retries times RTT,'' which is not a bound
in an asynchronous model, and a lower bound that fails whenever an
earlier independent indication exists; it is restated and split into a
proposition and a conditional corollary.  The characterisation of FLP's
proof was wrong and is corrected.

\emph{Withdrawn claims.}  That OAE ``eliminates'' ghosts (it bounds and
dates them; a residual \texttt{IN-DOUBT} state remains).  That ``no ghost
nodes can exist.''  That ``replace the channel model and the
impossibility dissolves'' --- a theorem does not dissolve when one
declines its assumptions.  \textbf{That a missing PIF echo identifies a
failed link} --- non-arrival is consistent with several causes, and
asserting otherwise would repeat the inference this paper criticises.
Any claim that OAE establishes \emph{common knowledge}.  Universals of
the form ``every mitigation,'' ``every interruption,'' ``every flap.''

\emph{Scoping.}  A new Section~\ref{sec:limitations} states hard
partitions, detection-is-not-repair, uncharacterised detector latency,
unproved triangle coverage, and the absence of any OAE measurement here.
Table~\ref{tab:mttf} is re-attributed to an industry slide, qualified at
first use, and carries the source's own definition of MTTF as
\emph{Mean Time to Flap}.  Silent data corruption and implicit failures
are explicitly identified as \emph{not} topology ghosts.  Two
architecture subsections outside the topology-knowledge argument are
scoped out rather than argued for.

\emph{The empirical survey of production systems is unchanged}, as is the
central claim: that delayed observation creates a knowledge interval, and
that making that interval explicit is different from guessing it away.}}
\end{center}

\vspace{0.5em}

\begin{abstract}
A link changes state faster than knowledge of that change can reach every
component that acts on it.  During the interval in between, some observer
operates on a topology that no longer exists.  We call that mismatch a
\emph{ghost}, and we define it precisely: for observer $i$, a ghost exists
whenever the connectivity state on which $i$ acts differs from the
connectivity actually available to it.  Ghosts appear wherever this
delay does---chiplet-to-chiplet (PCIe, UCIe), GPU-to-GPU (NVLink,
NVSwitch), node-to-node (Ethernet, Thunderbolt), and cluster-to-cluster
(IP, BGP).  What the mechanisms we survey have in common is narrower than
a shared ancestry: in each, a component's failure state is at some point
\emph{inferred} from absence, elapsed time, probing, or eventual
convergence, and absence alone does not distinguish delay from loss,
partition, or endpoint failure.

We survey the scale of the problem using production data from Meta
(419~interruptions in 54~days of LLaMA~3 training), ByteDance
(38,236~explicit and 5,948~implicit failures in three months), Google
(TPUv4 resiliency via optical circuit switching), and Alibaba (0.057\%
NIC--ToR link failures per month).  Under the assumptions of an industry
scaling projection we reproduce and attribute in Table~\ref{tab:mttf} ---
not our own measurement --- a fabric of ${>}10$~million optical links
would see an aggregate flap roughly every 48~seconds.  A flap creates a
ghost wherever knowledge of the prior topology outlives the physical
transition.

The mitigations we survey---Phi Accrual failure detectors, SWIM, BFD,
OSPF/ISIS fast convergence, SmartNIC offload, lossless Ethernet
(RoCE/PFC), and Kubernetes pod eviction---reduce the interval, often
substantially, but each retains a period in which failure state is
inferred rather than established.  We do not claim that no unilateral
mechanism can be adequate.  We connect
ghosts to two adjacent failure classes: \emph{gray failures} (Huang
et~al., HotOS~2017), where differential observability makes the ghost
invisible to monitoring, and \emph{metastable failures} (Bronson
et~al., HotOS~2021; validated empirically across 22~failures at
11~organizations, OSDI~2022), where retry amplification turns a single
ghost into a self-sustaining degraded equilibrium.  We further connect
link-layer ghosts to \emph{application-layer} ghosts via the companion
paper on AI/ML atomicity: non-atomic checkpoints
($\Pr[\neg\mathrm{Atomic}] \to 1$ as the number of shards grows) and
consensus-hard firmware deployment (the distributed firing-squad
problem).

We argue that Open \AE thernet (OAE) \emph{makes topology uncertainty
explicit and provides a protocol-defined mechanism for resolving it},
rather than converting elapsed time into a failure verdict, through a
Reliable Link Failure Detector (RLFD) operating at~L2, Perfect
Information Feedback (PIF), triangle failover, and atomic token
transfer.  Its distinctive move is asymmetric: a completed transaction
yields positive, mutually held evidence, while \emph{absence} of
completion is not treated as identification of a failure but recorded as
an explicit \texttt{IN-DOUBT} state to be resolved by corroboration
where the topology permits.  \textbf{We are explicit about what this
does not claim.}  OAE does not eliminate uncertainty; it makes it
first-class.  It does not repair a hard partition; detection is not
repair; its detector latency is uncharacterised; triangle coverage is
assumed rather than proved; we report no OAE measurement; and we do not
claim common knowledge in the epistemic-logic sense.
Section~\ref{sec:limitations} states these limits in full.
\end{abstract}

\clearpage
\tableofcontents
\newpage

%% =========================================================================
\section{Introduction: The Ghost in the Datacenter}
\label{sec:intro}

The largest AI training clusters in operation today contain millions of
GPUs connected by tens of millions of optical links arranged in
multi-tier Clos topologies.  These links flap.  The optical transceivers
that drive them have a mean time between hard failures (MTBF) measured in
millions of hours, but a mean time to \emph{flap} (MTTF) measured in
hundreds of thousands of hours.  At scale the distinction is decisive.
\textbf{Using the assumptions reproduced in Table~\ref{tab:mttf}} --- an
industry projection, not a measurement of ours --- a cluster with ten
million optics would experience a hard failure every few days and an
aggregate link flap roughly every 48~seconds.  We use that projection as
scale context throughout, and we do not independently validate it.

Every flap creates what we call a \emph{ghost}: a node, link, or path
that the network believes to be present but that is, in fact,
unreachable.  The ghost may be transient (the link recovers in
milliseconds) or persistent (the link enters a flap-suppress state or
degrades silently to a lower speed).  In either case, the network's
\emph{topology knowledge}---its understanding of its own graph---is
corrupted for some interval.  During that interval, traffic is routed to
a destination that cannot receive it, collective operations stall waiting
for a participant that will never respond, and higher-layer protocols
enter timeout-and-retry loops that amplify rather than resolve the
failure.

\begin{table}[ht]
\centering
\small
\begin{tabular}{lrrrr}
\toprule
\textbf{Year} & \textbf{Cluster (Power/GPUs)} & \textbf{Optics} & \textbf{Hard MTBF} & \textbf{Flap MTTF} \\
\midrule
2023 & 30\,MW / 20k GPUs    & ${>}100$k  & 4 days   & 3 hours    \\
2024 & 300\,MW / 200k GPUs  & ${>}1$M    & 7 days   & 12 min     \\
2025 & 4.5\,GW / 3M GPUs    & ${>}10$M   & 30 min   & \textbf{48 sec} \\
\bottomrule
\end{tabular}
\caption{Link failure scaling in AI training clusters.  Assumptions, as
stated on the source slide: 1500\,W/GPU, 3-layer Clos, one front-end
per four scaleout links; optics $10^7$\,hrs MTBF (hard), $3 \times
10^5$\,hrs MTTF (flap).  \textbf{Note the source's own definitions:}
MTBF is Mean Time Between Failures (hard), and \textbf{MTTF here
denotes Mean Time \emph{to Flap}}, not Mean Time To Failure.
\textbf{Source:} reproduced from the slide ``Bigger AI Clusters Need
Much Better Optics,'' presented at an Open Compute Project working
session, 6~November~2025; the originating organisation is being
confirmed and this table should not be attributed to the present author.
Previously reproduced in Borrill~\cite{borrill2026ocp}.}
\label{tab:mttf}
\end{table}

The ghost is not a new phenomenon.  It is the distributed-systems
manifestation of a structural error that has been embedded in networking
protocols since Shannon's 1948 formalization of the communication
channel~\cite{shannon1948}: the Forward-In-Time-Only (FITO)
assumption~\cite{borrill2026whatwrong}.  FITO treats every communication
link as a unidirectional, memoryless channel.  When a message is sent
and no acknowledgment arrives, the sender cannot know whether the message
was lost, the receiver crashed, or the link is simply slow.  The only
available response is to wait (timeout) and try again (retry).  This
Timeout And Retry (TAR) mechanism is the universal failure detector in
PCIe, NVLink, UALink, Ethernet, TCP, NCCL, Paxos, Raft, and
Kubernetes.

Fischer, Lynch, and Paterson~\cite{flp1985} proved in 1985 that no
deterministic protocol can guarantee consensus termination in an
asynchronous system with even one crash fault.  \textbf{We state the
relationship carefully, because it is easy to overstate.}  The core of
that proof is a bivalence argument about reachable configurations, not a
claim about link-state observation; slow/dead indistinguishability under
asynchrony is what makes the adversary argument available, rather than
being the result itself.  The topology problem considered here has a
similar \emph{epistemic} shape --- an observer cannot separate delay from
failure using absence alone --- but Proposition~\ref{prop:tar} does not
depend on FLP and is not weakened if the analogy is rejected.  Timeout
and retry is one common engineering response to that ambiguity; it does
not resolve it, but assigns a time after which ``slow'' is treated as
``dead.''

This paper traces the ghost from its origin in the FITO assumption
through its manifestation in PCIe (Section~\ref{sec:pcie}), NVLink and
NCCL (Section~\ref{sec:nvlink}), and datacenter-scale AI training
(Section~\ref{sec:scale}); surveys the failure of every existing
mitigation to eliminate it (Section~\ref{sec:solutions}); connects it
to the system-level failure modes of metastable failures and gray
failures (Section~\ref{sec:metastable}); links it to application-layer
ghosts via non-atomic checkpoints and consensus-hard firmware deployment
(Section~\ref{sec:atomicity}); documents its appearance in consumer
hardware (Section~\ref{sec:practice}); shows how Open \AE thernet (OAE)
makes the knowledge interval explicit at the link layer by replacing
timeout inference with transactional topology state
(Section~\ref{sec:oae}); and states the limits of
that claim (Section~\ref{sec:limitations}).

%% =========================================================================
\section{Unilateral Observation and the Knowledge Interval}
\label{sec:fito}

\subsection{Unilateral Observation and Timeout-Based Inference}

Shannon's 1948 mathematical theory of communication~\cite{shannon1948}
models a communication channel as a conditional probability distribution
$p(y|x)$ mapping input symbols to output symbols.  The channel is
memoryless: each transmission is independent.  The channel is
forward-in-time-only: information flows from sender to receiver, and the
receiver's only recourse upon detecting an error is to request
retransmission.  This model was appropriate for telegraph and telephone
circuits, where the physical medium was a dedicated point-to-point wire
and the failure mode was noise.

Every modern interconnect protocol inherits this model.  PCIe's Data
Link Layer retransmits packets that receive NACK or that timeout on
ACK/NAK.  NVLink's recovery mechanism re-trains the link and replays
failed transactions.  Ethernet's TCP retransmits segments that are not
acknowledged within a timeout.  NCCL's RAS subsystem detects stalled
collectives and restarts them.  In every case, the failure detector is a
\emph{timeout}, and the recovery mechanism is a \emph{retry}.

\subsection{Relation to Asynchronous Distributed-System Models}
\label{sec:relation}

\noindent\fbox{\parbox{0.95\linewidth}{\small \textbf{This subsection is
context, not load-bearing.}  Proposition~\ref{prop:tar}, the evidence in
Sections~\ref{sec:pcie}--\ref{sec:practice}, and the analysis of OAE in
Section~\ref{sec:oae} are all independent of everything argued here.  A
reader who rejects this framing entirely should find the rest of the
paper unaffected.  It is placed here because the question is asked often
enough that leaving it unaddressed would be its own kind of evasion.}}

\medskip

The observation that a system cannot separate ``slow'' from ``dead'' using
absence alone is not new, and it would be wrong to present it as such.  It
is the condition under which Fischer, Lynch and Paterson~\cite{flp1985}
prove that no deterministic protocol guarantees consensus termination, and
it is adjacent to the tradeoff Brewer~\cite{brewer2000} describes.  The
question worth asking is not whether those results are true --- they are
--- but what their \emph{antecedents} require, and whether a given piece
of hardware satisfies them.

FLP's antecedent is \emph{asynchrony}: unbounded message delay and
unbounded relative processing speed.  \textbf{A theorem does not dissolve
when one declines its assumptions}, and this paper makes no claim to
invalidate FLP, CAP, or the consensus hierarchy~\cite{herlihy1991}.  What
it observes is narrower: a link whose slice timing is physically bounded
at the SerDes layer is not an asynchronous system in FLP's sense, so
whether a particular fabric falls under the antecedent is an engineering
choice rather than a law of nature.  Establishing that a real fabric
satisfies a partial-synchrony model, and with what parameters, is work
this paper does not do.

Borrill~\cite{borrill2026whatwrong} develops the stronger position that
several impossibility \emph{intuitions} in distributed computing are
overgeneralised from models offering weaker interaction semantics than
current hardware supports, and calls the overgeneralisation a
\emph{category mistake} in Ryle's sense~\cite{ryle1949}: applying a
concept adequate for point-to-point data transfer to the different task
of maintaining shared knowledge of a graph.  \textbf{That is a design-space
critique, not a theorem}, and it is argued elsewhere rather than here.

\begin{definition}[Ghost]
\label{def:ghost}
A \emph{ghost} is a discrepancy between a network's topology knowledge
(the graph it believes it has) and its actual topology (the graph it
actually has).  Formally, let $G_K(t)$ be the topology graph known to
the network at time $t$, and let $G_A(t)$ be the actual topology graph.
A ghost exists whenever $G_K(t) \neq G_A(t)$.
\end{definition}

\begin{proposition}[Delayed observation creates a knowledge interval]
\label{prop:tar}
Let a connectivity predicate change at time $t_0$, and let observer $i$
continue to act on the pre-change value until it revises its state at
$t_d > t_0$.  Then $i$ has a topology ghost on $[t_0, t_d)$, of duration
$\Delta_g = t_d - t_0$.
\end{proposition}

\begin{corollary}[Timeout-only detectors]
\label{cor:timeout}
If the earliest mechanism by which $i$ can revise its state is expiration
of a timeout $T$, then for failures that produce no earlier independent
indication, $\Delta_g \ge T$.
\end{corollary}

\noindent Two things this deliberately does \emph{not} say.  It gives no
upper bound on $\Delta_g$: in an asynchronous model the round-trip time
is unbounded, so ``timeout plus retries times RTT'' is not a bound at
all.  And the lower bound is conditional, because many links do carry an
earlier independent indication --- loss of carrier, an AER event, a BFD
session drop --- in which case detection precedes the timeout and
$\Delta_g < T$.  \textbf{The proposition is a statement about delayed
observation, and it requires neither Shannon nor FLP.}  That is its
strength: it is nearly definitional, and correspondingly hard to
dispute.

\noindent\emph{Notation.} We write $\Delta_g$ for the \emph{ghost
window} under TAR.  This is deliberately distinguished from $\Delta$ as
used in the companion paper on partition surfaces, where it denotes the
OAE reconciliation bound; the two symbols refer to different quantities
and should not be conflated across the series.

In practice, $\Delta_g$ ranges from microseconds (PCIe completion
timeout, Range~A) to minutes (Kubernetes node-monitor-grace-period) to
hours (BGP route convergence under flapping suppression).  At every
timescale, the ghost exists.

%% =========================================================================
\section{PCIe: The Ghost Machine}
\label{sec:pcie}

PCIe is the foundational interconnect for modern compute
infrastructure.  Every GPU, NIC, NVMe drive, and accelerator connects
to its host through PCIe.  PCIe's link management protocol---the Link
Training and Status State Machine (LTSSM)---is also the prototype for
the ghost-creation mechanism that recurs at every scale.

\subsection{LTSSM and the Recovery State}

The LTSSM manages link negotiation through 11 top-level states: Detect,
Polling, Configuration, Recovery, L0 (normal operation), L0s, L1, L2,
Hot Reset, Loopback, and Disable~\cite{colton2024}.  The full LTSSM
traversal takes approximately 9\,ms, dominated by the Recovery state.

When a link experiences signal integrity issues, the LTSSM enters
Recovery.  If equalization succeeds within 2\,ms, the link returns to
L0.  If it fails, the LTSSM returns to Detect and restarts the entire
process.  During recovery, the link is neither up nor down---it is a
ghost.  Higher layers (Transaction Layer, software drivers) may or may
not have timed out on pending completions.  The result is ambiguous
topology state: some components of the system believe the device is
present; others do not.

\subsection{Completion Timeouts}

PCIe defines programmable completion timeout ranges~\cite{pcisig2022}:

\begin{table}[ht]
\centering
\small
\begin{tabular}{lr}
\toprule
\textbf{Range} & \textbf{Timeout} \\
\midrule
A        & 50--100\,$\mu$s \\
B        & 1--10\,ms \\
AB       & 50\,$\mu$s -- 10\,ms \\
ABC      & 50\,$\mu$s -- 50\,ms (default) \\
ABCD     & 50\,$\mu$s -- 64\,s \\
Flit mode & 40--50\,ms \\
\bottomrule
\end{tabular}
\caption{PCIe completion timeout ranges.  Each is an arbitrary choice;
none corresponds to a physical property of the link.}
\label{tab:pcie-timeouts}
\end{table}

A completion timeout does not mean the link is dead---it means the
requester has given up waiting.  The distinction between ``slow'' and
``dead'' is precisely the ambiguity that FLP proves is unresolvable.
PCIe papers over this with configurable timeouts, but the ambiguity
propagates upward through every layer of the stack.

\subsection{Silent Speed Degradation}

PCIe Gen5 (32\,GT/s) and Gen6 (64\,GT/s) face escalating signal
integrity challenges: inter-symbol interference increases with
frequency, equalization complexity grows, and parasitic capacitance and
stub resonances at Gen6 can shift return loss by more than
10\,dB~\cite{sigintegrity2025}.  When the PHY Rx equalization
computation exceeds the LTSSM timeout, the link silently falls back to
Gen1 (2.5\,GT/s)---a 12.8$\times$ bandwidth reduction---without
generating any Advanced Error Reporting (AER) error.  Training continues
at degraded performance with no notification.

This is a particularly insidious ghost: the link is ``up'' in every
observable sense, but its performance is 12.8$\times$ worse than
expected.  Detection requires periodic polling of PCIe link status
registers---polling for ghosts.

\subsection{AER and the Bus Reset Cascade}

PCIe's Advanced Error Reporting classifies errors as correctable (no
software intervention), uncorrectable non-fatal (transaction unreliable,
link functional), and uncorrectable fatal (link unreliable).  Fatal
errors trigger a Secondary Bus Reset, which affects \emph{all}
subordinate devices.  During reset, every device below the affected port
is a ghost---present in the operating system's device tree but
unreachable.  AER provides no mechanism for neighboring devices to learn
about the reset or update their topology knowledge.

\subsection{Forward Error Correction: Another Band-Aid}

PCIe 6.0 introduces lightweight LDPC forward error correction with a
target latency below 2\,ns~\cite{cadencefec}.  The 256-byte FLIT
structure allocates 242~bytes for payload, 8~bytes for CRC, and 6~bytes
for FEC, targeting a link retry probability below
$5 \times 10^{-6}$.  FEC reduces the symptom (bit errors) but does
not address the cause: the protocol has no transactional guarantee.
A corrected bit error and a silent link degradation are
indistinguishable from the FEC's perspective.

%% =========================================================================
\section{NVLink, NCCL, and the GPU Cluster}
\label{sec:nvlink}

\subsection{NCCL: Opaque Topology Construction}

NVIDIA's Collective Communications Library (NCCL) manages GPU-to-GPU
communication for distributed training.  A recent analysis
notes~\cite{demystifynccl2025}: ``Although NCCL is critical to
large-scale GPU systems and open source, its internal mechanisms remain
insufficiently documented\ldots\ key aspects such as \emph{topology
construction}, algorithm selection, pipelining, and buffer management
across nodes and devices are not clearly described.''

NCCL discovers topology automatically and selects communication
algorithms based on available interconnects.  When NVLink is available,
it takes precedence over PCIe.  But NCCL's topology discovery is a
snapshot taken at initialization time.  If a link fails after
initialization, NCCL's topology map becomes a ghost: it describes a
graph that no longer exists.

\subsection{NVLink Failure Modes}

NVLink failure modes include firmware version mismatches between GPUs
(causing cryptic collective operation failures), Fabric Manager crashes
on NVLink4 systems within one month of
operation, inband message errors on ports outside active partitions,
driver/Fabric Manager version mismatches that prevent NVSwitch fabric
availability entirely, and topology consistency bugs that lead to
unrecoverable depth-first-search states in NCCL's topology discovery.

\subsection{NCCL RAS: Timeouts All the Way Down}

Starting with NCCL~2.24, a Reliability, Availability, and
Serviceability (RAS) subsystem was added for
diagnosis~\cite{ncclras2024}.  It relies on multiple timeouts ranging
from 5 to 60~seconds---precisely the TAR approach.  The RAS subsystem
can detect that something went wrong, but it cannot distinguish ``slow
NVLink'' from ``dead NVLink'' from ``partitioned NVSwitch.''

NVLink tracks recovery counters (the number of times a link transitioned
to recovery and succeeded) and error count thresholds.  When thresholds
are exceeded, the link is marked degraded.  But silent performance
degradation does not generate XID errors---detection requires periodic
PCIe link status querying~\cite{ncclras2024}.  This is, once again,
polling for ghosts.

\subsection{NVIDIA Fabric Manager}

The Fabric Manager provides NVSwitch topology management: discovery,
local identifier assignment, forwarding table computation, and
monitoring~\cite{nvfm}.  But the Fabric Manager itself is a
centralized, polling-based system.  Failures are detected after the next
polling interval.  If the Fabric Manager crashes or becomes congested,
NVLink failures go unnoticed.  The Fabric Manager's control-plane
coupling means that a single process failure can mask all NVLink
failures from the rest of the system.

%% =========================================================================
\section{The Scale Crisis: Numbers from the Front Lines}
\label{sec:scale}

\subsection{Meta LLaMA 3 (2024)}

Meta's LLaMA~3 training run used 16,384 NVIDIA H100 80\,GB GPUs over
54~days~\cite{kokolis2024}.  The run experienced
\textbf{419~unexpected interruptions}---one every three hours.
Hardware failures caused 78\% of interruptions, with GPU failures
(including NVLink errors) at 30.1\% and HBM3 memory failures at 17.2\%.
Meta maintained ${>}90$\% effective training time through automated
recovery.  \textbf{We do not claim that all 419 were ghosts.}  The same
study attributes a large share to GPU and HBM faults, which are component
failures rather than topology-knowledge mismatches.  The subset that
bears on this paper is the connectivity and communication failures; the
aggregate figure is scale context, not a count of measured ghosts.

\subsection{ByteDance MegaScale and ByteRobust}

ByteDance's MegaScale system trains 175B-parameter transformers on
12,288~GPUs, achieving 55.2\% Model FLOPs
Utilization~\cite{megascale2024}.  Over a multi-week production run on
10,000+ GPUs, the system experienced more than 100~failure recovery
events.

ByteRobust~\cite{byterobust2025} achieved 97\% Effective Training Time
Ratio over three months on 9,600~GPUs, but detected
\textbf{38,236~explicit failures} and \textbf{5,948~implicit failures}
in that period.  \textbf{Implicit failures are not topology ghosts under
Definition~\ref{def:ghost}} --- they may be numerical, software, or
component faults --- but they are epistemically adjacent: the system
degrades without an explicit fault signal, so its operating model and its
actual condition diverge.
ByteRobust detects them by monitoring loss and gradient norm spikes
(5$\times$ increase), meaning the ghost was already corrupting training
before detection.

\subsection{Unicron: The Failure Pie Chart}

He et~al.~\cite{unicron2024} report task termination statistics from
large-scale LLM training.  Network-related failures---NCCL timeout
(10.1\%), link flapping (3.9\%), other network errors (4.0\%), and
connection refused (2.1\%)---total \textbf{20.1\%} of all task
terminations.  These are all manifestations of the same ghost: the
network's inability to know its own topology.

\subsection{Google TPUv4 Resiliency}

Zu et~al.~\cite{google_tpuv4_2024} describe Google's approach to
TPUv4 resiliency: 99.98\% system availability through optical circuit
switching (OCS).  When a failure is detected, the OCS fabric manager
optically bypasses unhealthy units.  This is reconfiguration rather than
timeout-and-retry, and it is among the most capable mitigations in
production.  But failure detection still relies on polling
(\texttt{healthd} monitoring and \texttt{libtpunet} status checks), and
the interval between failure occurrence and OCS reconfiguration is a
candidate ghost window.  \textbf{We do not equate that window with the
availability figure}: 99.98\% availability corresponds to roughly
105~minutes of unavailability per pod-year, but unavailability is the
\emph{consequence} of a failure, not a measure of how long stale
topology state persisted.  Deriving ghost duration would require
incident-level detection timestamps, which the cited work does not
report.

\subsection{Alibaba HPN}

Xi et~al.~\cite{alibaba_hpn2024} report from production: 0.057\% of
NIC--ToR links fail per month, and 0.051\% of ToR switches encounter
critical errors or crashes per month.  Under these rates, a single LLM
training job encounters 1--2 crashes monthly.  Alibaba's non-stacked
dual-ToR design eliminates ToR-related single-node failures---a
structural anti-ghost measure that reduces blast radius but does not
eliminate TAR-based detection.

\subsection{GPU Hardware: H100 vs.\ A100}

The H100 has 3.2$\times$ lower per-GPU mean time between errors (MTBE)
for memory errors compared to the A100~\cite{h100vsa100}.  NVLink
errors impede data transfer and reduce computational throughput but
cannot be handled by application-level
mechanisms~\cite{ampere_memory2025}.  A 3,000-GPU cluster has a peak
MTBF of 56.6~hours~\cite{ampere_memory2025}.  More powerful GPUs
produce more ghost-creating events per unit time.

\subsection{The Checkpoint Tax}

Non-atomic checkpointing consumes 12--43\% of training
time~\cite{borrill2026atomicity}.  Remote storage checkpoints take
30--40~minutes every three hours.  Meta's OPT-175B run (992~A100s,
60~GPU-days) required more than 105~restarts---one every
14~hours.  At current compute costs, hourly checkpointing on a
512-GPU cluster costs more than \$4,000 per day.  This is the tax for
not having atomic snapshots---a cost that the AI/ML Atomicity
paper~\cite{borrill2026atomicity} shows is structural, not merely an
engineering problem.

\subsection{Silent Data Corruption}

Silent data corruption (SDC) is \textbf{not} a topology ghost under
Definition~\ref{def:ghost}, and we flag that explicitly rather than
stretching the term.  It is a distinct form of latent state divergence,
included here because it shares the property that the system's model of
its own health outlives the truth.  The NVIDIA/OCP
whitepaper on SDC in AI~\cite{nvidia_sdc2025} documents that minor SDCs
result in biased or incorrect outputs, and that core-based anomaly
detection catches SDCs 41\% better than testing approaches.  Dixit
et~al.~\cite{dixit_sdc2021} tested hundreds of thousands of machines
and detected hundreds of CPUs with SDC defects.  A single persistent
SDC device cascades corruption through NCCL collectives to all nodes in
the training cluster.

%% =========================================================================
\section{Why Existing Solutions Still Create Ghosts}
\label{sec:solutions}

The mechanisms surveyed below reduce detection latency, blast radius, or
recovery time, and several are very effective at what they set out to do.
\textbf{What they have in common is narrower than a universal claim and
is the only thing this section asserts}: in each, some component's failure
state is at some point \emph{inferred} from absence, elapsed time,
probing, or eventual convergence, and during that inference a stale-state
interval exists for some observer.  We do not claim that no unilateral
mechanism can be adequate, nor that these mechanisms are alike in
anything but that respect.

\subsection{Failure Detectors}

The \emph{Phi Accrual} failure detector~\cite{hayashibara2004}
dynamically adjusts suspicion levels based on heartbeat inter-arrival
times, replacing binary up/down decisions with a continuous suspicion
score.  It is used in Apache Cassandra and Akka.  But it remains
timeout-based: the suspicion score is computed from \emph{absence} of
heartbeats, and FLP guarantees that this absence is ambiguous.

The \emph{SWIM} protocol~\cite{swim2002} separates failure detection
from membership dissemination using peer-to-peer randomized probing and
gossip-style epidemic propagation.  Network load is $O(\log n)$ per
node.  But SWIM provides only weak consistency: at any point, different
nodes have different views of membership---different ghosts.

\emph{BFD} (Bidirectional Forwarding Detection,
RFC~5880/5881~\cite{bfd2010}) can detect link failures in under
50\,ms.  But BFD enters suppression mode after three flaps in
15~seconds, backing off to higher intervals---defeating the purpose of
fast detection precisely when link flapping is most active.  And
50\,ms is still a timeout.

\emph{Kubernetes} uses a 40-second \texttt{node-monitor-grace-period}
before declaring a node unreachable, followed by a 300-second default
pod eviction timeout.  During this 40--340~second window, the kubelet on
the partitioned node continues running pods while the control plane
believes them dead.  This is a ghost that can last nearly six minutes.

\subsection{Routing Convergence}

OSPF and ISIS Loop-Free Alternates (LFA) pre-compute backup paths, and
Topology-Independent LFA (TI-LFA) with segment routing can achieve
local rerouting in under 50\,ms.  But global state must still propagate
via TAR.  During the convergence window, some routers have old routes
and some have new ones, creating micro-loops and black
holes---ghosts in the forwarding plane.

\subsection{Datacenter Network Architectures}

Google's Jupiter/Aquila~\cite{jupiter2022,aquila2022nsdi} uses SDN with
adaptive routing: failures manifest as congestion, and adaptive routing
works around them until the SDN engine removes failed entries.  But the
delay between failure occurrence and SDN reprogramming is the ghost
window.

Meta's F16 fabric~\cite{metaf16} uses 16 parallel 100\,GE planes
instead of 4$\times$400\,GE to provide redundancy.  But redundancy does
not eliminate detection: the fabric still uses BFD for link failure
detection.

Clos topologies provide path diversity, but correlated failures---power
events, code deployments, hardware batch defects---can affect multiple
components simultaneously, overwhelming the path diversity.

\subsection{Hardware Acceleration}

SmartNICs and DPUs (NVIDIA BlueField, Intel IPU) offload failure
detection to dedicated hardware, reducing OS scheduler variability.  But
moving the timeout to hardware does not eliminate the timeout.  Worse,
the SmartNIC itself can fail, creating a new class of ghost: the host
and the SmartNIC disagree about system health.

\subsection{Lossless Ethernet (RoCE)}

RoCE uses Priority Flow Control (PFC) to make Ethernet lossless for
RDMA~\cite{dcqcn_deadlocks}.  But PFC creates deadlocks: circular pause
dependencies where no packets in the cycle can move even when congestion
clears.  PFC Watchdog catches pauses exceeding 200\,ms---another
timeout---and breaks the deadlock by \emph{dropping} packets, reverting
to lossy behavior.  Making Ethernet ``lossless'' with PFC merely
transforms packet-loss ghosts into deadlock ghosts.

R-Pingmesh~\cite{rpingmesh2024}, deployed on tens of thousands of RNICs
for six months, detects RNIC flapping, link flapping, and fiber damage
with 85\% accuracy.  Hostmesh~\cite{hostmesh2024} found that hardware
failures comprise the majority of RoCE network problems, with RNIC
flapping, switch link flapping, and packet corruption as the most common
failure modes.

\subsection{The Gray Failure Problem}

Huang et~al.~\cite{grayf2017} define \emph{gray failure} as a partial
failure with differential observability: some components see the system
as healthy while others see it as failed.  Gray failure is the ghost's
natural habitat.  A node that passes all BFD checks, responds to pings,
and returns HTTP~200 from its health endpoint can still be partially
failed---one CPU core faulty, NVLink degraded, memory thrashing.  The
application sees catastrophically slow responses while the
infrastructure sees a healthy node.

The fundamental issue: failure detection cannot be solved at the failure
detection layer because the problem is \emph{differential
observability}---different parts of the system have different
information about failures, and no TAR mechanism can synchronize this
information in bounded time across an asynchronous network.

%% =========================================================================
\section{Metastable Failures: The Ghost That Sustains Itself}
\label{sec:metastable}

Bronson et~al.~\cite{bronson2021} define a \emph{metastable failure} as
a system state where a trigger causes transition to a degraded
equilibrium that persists even after the trigger is removed.  Two
components are required: a \emph{trigger} (load spike or capacity
decrease) and a \emph{sustaining effect} (a positive feedback loop that
amplifies the degradation).

Huang et~al.~\cite{huang_metastable2022} validated the concept
empirically, studying 22~metastable failures from 11~organizations
(AWS, Google, Azure, IBM, Spotify, and others).  Key findings:

\begin{itemize}
\item At least 4 of 15 major AWS outages in the last decade were
  metastable failures.
\item Outage durations ranged from 1.5 to 73.53~hours.
\item \textbf{Retry amplification} was the most common sustaining
  effect, present in more than 50\% of all incidents.
\item Engineer errors triggered ${\sim}45$\% of incidents; load spikes
  triggered ${\sim}35$\%.
\end{itemize}

The retry amplification mechanism is quantifiable.  A system nominally
handling 300~QPS becomes overloaded; latency exceeds the timeout
threshold; clients retry; load rises to 560~QPS (280~original plus
280~retries); the system cannot recover even when the original trigger
is removed.  RetryGuard~\cite{retryguard2025} reports that default retry
handling can drive resource consumption to more than 1000\% of baseline.

Production examples include the Slack outage of February~2022 (cache hit
rate drops $\to$ increased database load $\to$ database overload $\to$
cache cannot rebuild $\to$ 73.53-hour cascading failure); the AWS
DynamoDB DNS race condition of October~2025 (cascaded through
141~AWS services); and the Azure networking configuration change of
February~2025 (${\sim}50$-hour outage with cascading retry storms).

\subsection{Why Metastable Failures Are Ghosts}

The metastable failure is the ghost that \emph{feeds on itself}.  The
system is simultaneously ``up'' (accepting connections, processing
requests) and ``down'' (delivering zero useful throughput).  Monitoring
sees a healthy system; users experience total failure.  This is gray
failure (differential observability) plus positive feedback
(self-sustaining degradation).

The FITO connection is direct: every metastable failure in the OSDI~2022
dataset involves TAR.  The retry is the sustaining mechanism, and the
retry exists because the protocol uses timeout-based failure detection.
The ghost creates retries; retries create more ghosts; the ghosts
sustain each other.

\subsection{Mitigations and Their Limits}

Load shedding (used in ${>}55$\% of documented recovery cases) breaks the
feedback loop by dropping requests, but requires external
intervention---the system cannot shed load on its own once trapped.
Circuit breakers reject outgoing traffic to troubled services, but the
circuit breaker itself uses timeouts to decide when to open or close.
Capacity headroom (operating below the ``vulnerable'' threshold) is
economically infeasible for frontier AI training clusters, which are
designed to maximize utilization.

Isaacs et~al.~\cite{isaacs2025} propose predictive analysis using
queueing-theoretic models and discrete event simulators, but note that
the fundamental challenge remains: breaking the feedback loop requires
information about the system's state that TAR cannot provide.

%% =========================================================================
\section{The Atomicity Connection}
\label{sec:atomicity}

The companion paper, ``Why Atomicity Matters to AI/ML
Infrastructure''~\cite{borrill2026atomicity}, proves two results that
connect directly to the ghost.

\subsection{Non-Atomic Checkpoints}

Training state $S = (W, O, G, R, D)$---weights, optimizer state,
gradient buffers, RNG state, data iterator---spans GPUs, host memory,
NICs, NVMe, object stores, and metadata services.  A checkpoint is a
multi-step protocol execution, not a single event.  If each of $K$
atomic units persists with independent probability~$q$, then:
\begin{equation}
\Pr[\neg\mathrm{Atomic}(e)] = 1 - q^K - (1-q)^K
\label{eq:non-atomic}
\end{equation}
For any fixed $q \in (0,1)$, this probability approaches~1
exponentially fast as $K$ grows.  At frontier scale ($K$ in the
thousands), non-atomic checkpoints are the expected outcome.

Every non-atomic checkpoint creates an \emph{epistemic ghost}: a state
that looks valid but contains a causality violation.  If weights are
from epoch~$e$ but optimizer state from epoch~$e{-}1$, the update
$W_{e+1} = W_e - \eta O_{e-1}$ is not SGD.  The corruption manifests
as delayed divergence, mode collapse, or silent quality degradation.

\subsection{Firmware Deployment as a Firing Squad Problem}

Atomic fleet update---transitioning all XPUs from firmware $F_0$ to
$F_1$ without any mixed-protocol execution---is equivalent to the
distributed firing-squad
problem~\cite{coan1989,moses1988}.  The consensus
hierarchy~\cite{herlihy1991} guarantees that read/write messaging
(consensus number~1) cannot achieve agreement on the transition epoch
under failure.

During a firmware rollout, every node running $F_0$ while neighbors run
$F_1$ is a \emph{semantic ghost}---it appears functional but its
protocol semantics do not match the cluster's.  Mixed NCCL collectives,
mixed RDMA semantics, or mixed NVLink protocols can corrupt optimization
without any visible error.

\subsection{The Retry Amplification Loop}

The AI/ML Atomicity paper identifies a positive feedback loop:
checkpoint-and-restart increases load $\to$ load increases failure
probability $\to$ failures trigger more checkpoints.  This is the
ghost-creation machine in feedback mode---structurally identical to the
metastable failure mechanism of Section~\ref{sec:metastable}.

%% =========================================================================
\section{Ghost Nodes in Practice}
\label{sec:practice}

\subsection{EXO Labs: Ghost Nodes in the Visualization}

EXO is a distributed AI inference framework that connects consumer
devices into a cluster using pipeline-parallel inference, splitting
LLMs into ``shards'' assigned to different
devices~\cite{exo}.  EXO's topology visualization shows
the ghost problem directly: nodes appear in the display but are
unreachable, and when a Thunderbolt link drops, the node remains in the
topology map while inference shards sent to it are lost.  The
coordinator has no deterministic mechanism to distinguish ``slow node''
from ``dead node.''

\subsection{Thunderbolt on Mac Mini Fleet}

Thunderbolt~4 provides 40\,Gbps bidirectional bandwidth;
Thunderbolt~5 (M4~Pro/Max) provides 80\,Gbps.  Known link stability
issues on Mac Mini fleets include: Apple prohibits cycles in
Thunderbolt topology~\cite{thunderbolt_apple}; all Thunderbolt
connections show green (connected) but most ping connections fail
(request timeouts)---a classic ghost; every node reboot causes
Thunderbolt network interfaces to go down; and macOS software updates
can break USB/Thunderbolt networking at the kernel module
level~\cite{thunderbolt_webai,exo_issue525}.

The Mac Mini fleet demonstrates the same ghost problem seen at
datacenter scale, made visible on consumer hardware---and is the
platform on which the OAE protocol \emph{is being implemented and
evaluated}, with the intent of detecting link disconnections
deterministically and repairing graph infrastructure knowledge after
link events.  \textbf{We report no OAE measurement from this platform
in the present paper}; the rig, its fault-injection method and its
results are the subject of separate work.

\subsection{Network Partition Data}

Alquraan et~al.~\cite{alquraan2018} analyzed 136~failures from
25~widely-used distributed systems attributed to network partitioning.
The majority led to catastrophic effects: data loss, deleted data
reappearance, broken locks, and system crashes.  Their NEAT testing
framework found 32~new partition-related failures in 7~popular systems.

Kingsbury's Jepsen project~\cite{kingsbury_jepsen} systematically tests
distributed databases under network partitions and repeatedly
demonstrates that systems claiming consistency guarantees fail
catastrophically when links flap.

\subsection{Rail-Optimized Topologies and Failure Amplification}

Rail-optimized topologies~\cite{railonly2023,railx2025} reduce network
cost by 38--77\% by reducing path diversity.  But reduced path diversity
means every link failure has larger impact: a ghost node in a rail
topology can partition an entire rail group, affecting all GPUs that
share that rail.  The cost savings of rail-only topologies make the
ghost problem worse, not better.

%% =========================================================================
\section{OAE: Making the Knowledge Interval Explicit}
\label{sec:oae}

Open \AE thernet (OAE)~\cite{borrill2026ocp,oaespec} \emph{makes the
knowledge interval explicit} at the link layer by replacing timeout
inference with transactional
topology knowledge.  The claim is that uncertainty becomes an explicit
protocol state rather than an implicit consequence of elapsed time;
Section~\ref{sec:limitations} says where it stops.

\subsection{Reliable Link Failure Detector (RLFD)}

The RLFD operates at L2 (below the transport layer), embedded directly
into chiplet fabrics.  Where a link forms part of a three-node
consensus group (a \emph{triangle}), link status can be verified
deterministically rather than by waiting for an arbitrary timeout: the
triangle detects the failure as a consequence of the interaction
protocol itself, not after a timeout and not after a retry.

\paragraph{Two limits, stated here rather than left to the reader.}
\emph{Latency.} ``Deterministic'' is a statement about the
\emph{mechanism}, not about elapsed time.  \textbf{We do not yet report
a latency bound for the RLFD.}  The detection interval is a function of
slice duration and the number of protocol rounds required to reach
agreement, and characterising it---together with the constant $k$ left
open in the specification---is open work.  Until that number exists, the
honest claim is that RLFD replaces an \emph{unbounded} guess with a
\emph{protocol-defined detection whose worst-case latency remains to be
characterised}.  We deliberately avoid the word \emph{bounded}: the
protocol supplies a finite state and round structure, but this paper
proves no wall-clock bound under an explicit timing model, and no
quantitative bounded-detection claim is made here.
\emph{Coverage.} Triangle membership is an assumption about topology,
not a theorem.  In an arbitrary graph, and at the boundary of a finite
planar mesh in particular, some links have restricted neighbourhoods and
lie in no triangle.  For those links the mechanism degrades: PIF still
gives the two endpoints a common view of what crossed the wire, so a
\emph{one-sided} detection remains available, but the third-party
corroboration that makes the detection deterministic does not, and such
links fall back to the residual \texttt{IN-DOUBT} state described below.
\textbf{We do not prove triangle coverage for arbitrary OAE topologies,
and a deployment that requires deterministic detection on every link
must construct a topology in which every link is triangle-covered.}

\subsection{Perfect Information Feedback (PIF)}

Every slice of data transmitted over an OAE link is echoed in reverse
via a reversible link state machine.  PIF guarantees that both sides of
a link share a common view of what was transmitted and what was
received.

\paragraph{Positive and negative evidence are not symmetric, and this is
the substance of the mechanism.}  When a PIF transaction \emph{completes},
both endpoints hold positive, transactional evidence that the slice
crossed and was witnessed; that is knowledge, not inference.  When the
echo \emph{does not} arrive, the situation is materially different.
Non-arrival is consistent with a delayed echo, a failure in the reverse
direction, a failure of the remote endpoint, a fault in either state
machine, or a partition --- and \textbf{absent additional progress
assumptions or third-party evidence, non-arrival does not by itself
identify which.}  It would be inconsistent for this paper to criticise
timeout-based detection for reading absence as identification and then
have PIF perform the same inference in the opposite direction.  We
therefore claim only the positive half: \textbf{PIF makes successful
transfer mutually evident; it does not make failure self-identifying.}
Unresolved cases enter \texttt{IN-DOUBT} (Section~\ref{sec:oae}) and are
resolved, where the topology permits, by corroboration rather than by
elapsed time.

\paragraph{On common knowledge.}  We claim mutual evidence of a completed
transfer between the two endpoints.  \textbf{We do not claim common
knowledge in the epistemic-logic sense}; the coordinated-attack result
establishes that finitely many acknowledgments over an unreliable channel
do not produce it, and nothing in PIF evades that.  Earlier work in this
programme that asserted common knowledge as a foundation is withdrawn.

\subsection{Triangle Failover}

Every triangle in the OAE topology is a recovery resource.  When a link
fails, the triangle provides error detection and recovery without TAR.
The two surviving links of the triangle carry recovery traffic while the
graph is updated.  This is not redundancy (having a backup path); it is
\emph{structural recovery} (the topology itself contains the recovery
mechanism).

\subsection{Atomic Token Transfer}

The OAE specification defines an atomic token-transfer primitive
\emph{at the level of an individual link}, and proposes composition of
those primitives over larger structures --- chains, DAGs and trees ---
with the graph as the first-class object.  \textbf{We flag the size of
that second claim rather than assuming it.}  Atomicity of pairwise
transfers does not by itself confer atomicity on a graph operation
spanning many links: that requires a stated linearization point, a
consistency model for the distributed graph state, and a specified
behaviour when a participant fails mid-composition.  None of those is
established here, and the composition properties remain a subject for
formal analysis.  What follows describes the intended sequence when a
link event occurs:

\begin{enumerate}
\item Where the link is triangle-covered, the triangle detects the event
      deterministically---by protocol, not by timeout.
\item The link-state transition is bound to an explicit graph-state
      update.  \textbf{!}~\emph{Atomic with respect to whom} is exactly the
      question left open above: the consistency semantics of that update
      across multiple observers are specified in the architecture, not
      established here.
\item Routing tables are recomputed.
\item Ghost nodes are \emph{detected and dated} rather than inferred:
      topology knowledge is transactional, so the interval of
      uncertainty has a defined beginning and a defined end instead of
      running until a timeout expires.
\end{enumerate}

\noindent What remains is the residue.  A link that cannot be resolved
by its own endpoints and has no triangle to corroborate it enters an
explicit \texttt{IN-DOUBT} state, and stays there until a third party
resolves it or the link is quarantined.  \textbf{That state is a
knowledge gap, and OAE does not close it}---the difference from TAR is
that the gap is \emph{declared} rather than silently guessed at.  We
regard making the residue explicit as the substantive claim; ``no ghost
can exist'' would be a stronger claim than the specification supports.

\subsection{Out of scope for this paper}

The wider OAE architecture also proposes graph-derived configuration in
place of human-mediated configuration, and a frame format in which the
representation on the wire matches the representation in memory.  Neither
bears on the topology-knowledge claim evaluated here, and neither is
argued for in this paper; they are noted only so that a reader
encountering them in the specification does not take their absence here
for an omission.  \textbf{Nothing in Sections~\ref{sec:oae}
or~\ref{sec:unified} depends on them.}

%% =========================================================================
\section{From Chiplets to Clusters: Unified Semantics}
\label{sec:unified}

OAE works at every scale---chiplet-to-chiplet, GPU-to-GPU,
node-to-node, cluster-to-cluster---because the fundamental primitive
(atomic link transaction) is the same at every scale.  The ghost is
bounded when topology knowledge is transactional.

The bisynchronous FIFO literature~\cite{borrill2026bisync} provides the
silicon-proven existence proof: interaction-based synchronization
primitives (handshakes, mutual exclusion, causal flow control) have
operated correctly across independent clock domains for over four
decades without timestamps, shared clocks, or any form of FITO
assumption.  OAE extends this principle from on-chip clock-domain
crossings to network-scale topology management.

Rail-optimized topologies, which reduce cost by reducing path diversity,
make the ghost problem worse under TAR-based protocols (every link
failure has larger blast radius).  Under OAE, failures in rail
topologies are \emph{detected} deterministically wherever the link is
triangle-covered---but detection is not repair.  Reduced path diversity
is precisely what local repair consumes, so a rail topology can leave a
detected failure unrepairable: \textbf{repairability is a property of
the topology's cuts, not of the detector}.  The honest claim is that
OAE removes the ambiguity about \emph{whether} the link failed, and
leaves the routing consequences of a sparse topology exactly where it
found them.

%% =========================================================================
\section{Limitations and Threats to Validity}
\label{sec:limitations}

The argument of this paper is that TAR-based detection produces ghosts
as a structural consequence, and that a bilateral, transactional link
layer bounds them.  We think the first half is strong.  The second half
is a claim about a specification and an implementation in progress, and
this section states plainly where it stops.  \textbf{Throughout, the
normative artifact is the OAE specification~\cite{oaespec}; where this
paper describes intent and the specification describes mechanism, the
specification governs.}

\paragraph{OAE does not repair a hard partition.}
If the graph is genuinely cut, no link-layer protocol reconnects it.
OAE changes what the surviving components \emph{know}---each side can
determine that the link is down rather than guessing---but a bounded,
correctly-dated report of a partition is still a partition.  Claims in
this paper about recovery are claims about topology knowledge, not about
connectivity.

\paragraph{Detection is not repair.}
Following from the above: deterministic detection on a topology with
insufficient path diversity yields a failure that is known and still
unroutable.  Repairability is a property of the graph's cuts.  This is
the substance of the rail-topology discussion in
Section~\ref{sec:unified}, and it generalises.

\paragraph{The detector's latency is uncharacterised.}
We give no nanosecond, microsecond or round-count bound for the RLFD.
The interval is a function of slice duration and the rounds needed to
reach agreement, and the relevant constant is left open in the
specification.  For a paper whose thesis is that unbounded timeouts are
the problem, this is the most uncomfortable gap in it, and we prefer to
name it than to let ``immediately'' stand in for a number.

\paragraph{Triangle coverage is assumed, not proved.}
Deterministic detection depends on each link belonging to a three-node
consensus group.  We do not prove that every link in an arbitrary OAE
topology is triangle-covered, and at the boundary of a finite planar
mesh some are demonstrably not.  Uncovered links retain one-sided PIF
detection and fall back to \texttt{IN-DOUBT}.

\paragraph{The residue is real.}
\texttt{IN-DOUBT} is a knowledge gap that persists until a third party
resolves it, or indefinitely under quarantine.  OAE makes it explicit;
it does not make it empty.  A reader who takes ``no ghost can exist''
from an earlier version of this work should take this paragraph instead.

\paragraph{We report no OAE measurement.}
The Mac Mini/Thunderbolt platform of Section~\ref{sec:practice} is an
implementation and evaluation vehicle, not a result.  No detection
latency, no repair time and no failure-injection campaign is reported
here.  The empirical claims in this paper are all about \emph{existing}
systems---PCIe, NVLink, NCCL, and the published production incidents of
Section~\ref{sec:scale}---and those stand on their own citations.

\paragraph{One table is not ours.}
The scaling figures in Table~\ref{tab:mttf} are reproduced from a slide
presented at an OCP working session in November~2025.  We use them as a
published industry projection, not as our own measurement, and the
originating organisation is being confirmed; see the caption.

\paragraph{Scope of the FITO claim.}
That TAR is a design choice rather than a law is an argument about
inherited assumptions, not a proof that no unilateral protocol can be
adequate in practice.  Many systems tolerate ghosts perfectly well
because their applications are insensitive to the ambiguity.  Our claim
is that the cost is paid, is now measurable at scale, and is rarely
attributed to its cause---not that every deployment should pay to remove
it.

%% =========================================================================
\section{Conclusion}
\label{sec:conclusion}

The ghost in the datacenter is not a bug.  It is a consequence of the
definitions.  More precisely: wherever an observer continues to act on
pre-failure topology state until some later revision, a nonzero
knowledge interval exists, and that follows from delayed observation
alone (Proposition~\ref{prop:tar}).  It does not depend on FLP, and it
does not depend on any claim about Shannon.

Any protocol whose earliest revision of link state is the expiry of a
timeout will produce ghosts---nodes, links, and paths that some observer
believes to be present but that are, in fact, unreachable.  That follows
from delayed observation and the detector's own semantics.  The wider
claim that this pattern is inherited from a particular channel model, and
its relation to the asynchronous models in which FLP and CAP are stated,
is discussed in Section~\ref{sec:fito} as \emph{context}; nothing in
Proposition~\ref{prop:tar} or in the evidence of
Sections~\ref{sec:pcie}--\ref{sec:practice} rests on it.

At 2025 cluster scale, a link flap creates a ghost somewhere in the
fabric roughly every 48~seconds under the Table~\ref{tab:mttf}
projection.  Ghosts can contribute to gray failures (differential
observability makes them invisible to monitoring) and metastable
failures (retry amplification turns them into self-sustaining degraded
equilibria that persist for hours even after the trigger is removed).
Application-layer ghosts---non-atomic checkpoints and mixed-protocol
firmware deployments---compound the damage with epistemic corruption
that is invisible to the training loop.

The operational costs of recovery are substantial and well documented:
checkpointing consumes 12--43\% of training time in the cited studies,
spare capacity is provisioned at around 5\%, retry can amplify rather
than resolve, and human operators remain in the diagnosis path.
\textbf{We do not attempt to attribute those costs quantitatively to
topology ghosts}, and we are aware that most of them have several causes.
They establish that failure recovery is economically material, which is a
weaker and more defensible claim than the one this paragraph previously
made.

Open \AE thernet makes the knowledge interval explicit at the link
layer.  \textbf{We do not claim a bound}: no wall-clock detection bound
is established in this paper, and calling the interval bounded while
leaving the bound unproved would be the same move we criticise
elsewhere.
Deterministic failure detection (RLFD), Perfect Information Feedback,
triangle failover, and atomic token transfer make topology knowledge
transactional.  Because the ghost is the trigger for both gray failures
and metastable failures, bounding its duration removes the open-ended
trigger that allows these system-level failure modes to form.

The claim is not that uncertainty disappears.  It is that uncertainty
acquires a beginning, an end, and a name---and that a declared
\texttt{IN-DOUBT} state a system can act on is a different object from
a silent guess it cannot distinguish from the truth.  Whether the
remaining residue is small enough to matter in production is an
empirical question this paper does not settle.

%% =========================================================================
\section*{Acknowledgments}
%% =========================================================================

This paper was developed with AI assistance (Claude, Anthropic) for
literature organization, drafting, and \LaTeX{} preparation.  The
research program originates from the author's work on
hardware-verified deterministic networking (VERITAS, 2005), replicated
state machines (REPLICUS, 2008), atomic Ethernet (Earth Computing,
2012--2020), and the FITO category mistake framework (D{\AE}D{\AE}LUS,
2024--present).  All technical claims, interpretations, and arguments
are the sole responsibility of the author.

%% =========================================================================
%% References
%% =========================================================================

\end{document}